\documentstyle[12pt]{article}
\begin{document}
\begin{center}
{\large\bf BTW model on dilute lattice}
\end{center}
\vspace {0.4 cm}
\begin{center} {\large Ajanta Bhowal Acharyya} \end{center}
\vspace {0.3 cm}
\begin{center} {\it Department of Physics, Lady Brabourne College,}\\ 
{\it P-1/2 Suhrawardy Avenue, Calcutta-700017, INDIA}\\
{E-mail:ajanta.bhowal@gmail.com}
\end{center}
\vspace {1.0 cm}
\noindent {\bf Abstract:} 
The variation of $\bar z$ in BTW model in presence of holes (dissipative sites) has been
studied. The value of $\bar z$ decreases as the fraction of number of holes increases.
Interstingly, it is observed that the variation of the rate of change of $\bar z$
with the fraction of number of holes is different  for the two different 
types of distribution of holes over the lattice. When the holes are randomly distributed over 
the lattice then the dissipation is more compared to that of the case when the holes
are present in the form of a single compact cluster with same fraction. The value of $\bar z$ is less
in the first case than that observed in the second case.

\vspace {1 cm}
\leftline {\bf PACS Numbers: 05.50 +q}
\leftline {\bf Keywords: BTW model, SOC}

\vspace {3cm}
\leftline {------------------------------------}
\leftline {E-mail:ajanta.bhowal@gmail.com}
\newpage

\leftline {\bf I. Introduction}

There exists some extended driven dissipative systems in nature which
show self-organised criticality (SOC). This phenomena of SOC is
characterised by spontaneous evolution into a steady state which shows
long-range spatial and temporal correlations.
 The simple lattice automata model of sandpile which shows
this SOC behaviour was introduced by 
Bak, Tang and Wiesenfeld (BTW)\cite{btw}.
The steady state dynamics of the model shows a power law in the probability
distributions for the occurence of the relaxation (avalanches) clusters of a
certain size, area, lifetime, etc. Extensive work has been done so far to
study the properties of the model in the steady SOC state\cite{dd,evi,vbp,bb,lub,ev,piet,ab}. 


Dissipative sandpile model has also been studied recently. However, it is in
question whether the SOC state is reached by dissipative model. Manna, Kiss and Kertesz \cite{mkk} 
studied a sandpile mdel in presence of dissipation. In this model, the dissipation of a
grain is considered during a relaxing event in a probabilistic manner. The numerical results
showed that the system reaches a subcritical state, with a characteristic size of the avalanches
depending inversely on the probability of dissipation. On the other hand, Manna et al. \cite{manna2} 
studied the directed dissipative sandpile model in two dimensions with observation that the
long time steady state is critical. Malcal et al \cite{malcal} studied the dissipative sandpile
model (with closed boundary) and found that the scaling properties are in the universality
class of the stochastic Abelian model with conservative dynamics and open boundaries. Najafl
et al \cite{najafl} studied the statistics of avalanches and the wave frontiers in dissipative Abelian sandpile
model.

In this paper, we have studied the average value of the height variable (averaged
over all the lattice sites) on the dilute lattice (in presence of some holes
or dissipative sites on the lattice), in steady state.
 We have considered the holes on the lattice created in two
 different ways: (i) the holes are created  at randomly selected sites of the lattice and
(ii) the cluster of holes are in the form a large cluster of square shape,
 situated around the central site of the lattice.
In this paper, we have studied how the value of average height variables 
(${\bar z}$)
decreases as the fraction of number of holes increases, in both the cases. 


\bigskip
\leftline {\bf II. The model and simulation}

The BTW model is a lattice automata model which shows some important properties
of the dynamics of the system, evolving  spontaneously towards a critical state. 
We consider a two dimensional square lattice of size
$L \times L$ with some holes on it.
 Out of these $L^2$ site, a particular
fraction of sites are chosen randomly  as hole. If any particle falls on any 
one of these holes, the particle will dissipate through the hole.

The model can be described in the following way: At each site $(i,j)$
of the lattice, a variable (so called height) $z(i,j)$ is associated
which can take positive integer values.
The height variable associated with the hole-sites  
remains zero for all time.  In every time step, one particle
is added to a randomly chosen site according to
\begin{equation}
z(i,j)=z(i,j)+1. 
\end{equation}

\noindent If, at any site  the height variable exceeds a critical value
$z_m$ (i.e., if $z(i,j) \geq z_m$), then that site becomes unstable and
it relaxes by a toppling. As an unstable site
topples, the value of the height variable, of that site is decreased by 4
units and that, of each of the four of its neighbouring sites  increased   
by unity (local conservation), i.e.,
\begin{equation}
z(i,j)=z(i,j)-4
\end{equation}
\begin{equation}
z(i, j \pm 1) = z(i , j \pm 1) +1 ~~{\rm and}~~
z(i \pm 1,j) = z(i\pm 1 ,j ) +1 
\end{equation}
\noindent for $z(i,j) \geq z_m$. Each boundary site is attached  to
an additional site which acts as a sink. We use here the 
open boundary conditions  so that the system can also 
dissipate  through the boundary. In our simulation, we have
taken $z_m = 4$. In this paper, the system can not only dissipate through
the boundary but also through holes present on the lattice. 

Now, we allow the system to evolve under the BTW dynamics (following eqns 1-3)
starting from an initial condition with all the sites having $z=0$.
As in the case of BTW model, here also the average height variable first
 increases  and ultimately it reaches  a steady  value.

 In this paper, we have
studied how this steady value changes  with the fraction of holes($f_h$) in two
 different cases of the arrangement of holes on the lattice. We have considered
two different arrangement of holes on the lattice, (i) The holes are distributed
 randomly on the lattice and (ii) the holes are in the form of a regular cluster 
of square shape around the central site of the lattice.

\bigskip
\leftline {\bf III. Results}

In our simulation, we first studied the variation of 
average height variable ($\bar z$) with time  for different values of the 
fraction of holes.
Fig.1 shows this variation when the holes are situated randomly on the lattice,
for a fixed system size ($L = 101$) for three different values of the
fraction of holes (0 to .2),
 for a particular configuration of the holes.
Fig.2 shows this variation when the holes form a cluster of square shape and 
placed at the centre of the square lattice of size ($L=101$). 
In this case we consider the size of the square shaped cluster of hole
as $L_c\times L_c$, where $L_c=(2*n_h+1)$ ($n_h$=1,11,21 )

From Fig.1 and Fig.2, it is observed that the 
value of the average height variable ($\bar z$) in the steady state decreases
as the fraction of holes ($f_h$) in the lattice increases,
in both the arrangement of holes.

We also studied the distribution of avalanche size in BTW model with holes(randomly 
scattered) for $L=101$, for $201$ different sample. And Log-Log plot of this distribution 
is plotted in the Fig. 3 together with that for BTW model. 

When the holes are created, randomly on the lattice, with 
a finite probability, then there will be a small
fluctuation in the number of holes ($N_h$) created.
In our simulation, for a fixed system size ($L = 101$),
for a particular probability of holes on the lattice,
we took the sample average of $N_h$ from $100$  different samples. 
And the average value of fraction of holes is caluculated using
the relation $f_h={N_h^2\over L^2}$.
We  also calculate the sample average of time average value of average height
 variable ($\bar z$) .  
We have taken the sample average of the fraction of holes($f_h)$ as well as 
time average of average value of height variable ($<\bar z>$). 

Whereas in the second case, when the holes are in the form of a square shaped
 cluster, we have taken the time average of the average value of height 
variable,
 for a particular size of the cluster (i.e, for a particular fraction of the
 holes in the cluster). The variation of average value of
height variables with the fraction of holes is studied in both the  cases.

It is interesting that the value of average height variable decreases with
 different rate as the fraction of holes increases  in the two cases.
Fig4 shows the variation of the value of average height variable with the
 fraction of holes for two different arrangement of holes  for $L=101$.

\vskip 1cm

\leftline {\bf IV. Summary}

The self-organised criticality is studied in a dilute square lattice by using
lattice automata (BTW) model. The dilution in the lattice is created in two
different ways. The (i) randomly distributed and (ii) regular clustered with
same fraction of dilution. The diluted site retains the value zero of the
BTW automaton throughout the dynamics towards the self organisation of 
critical state. Hence, the dilution acts like the points of dissipation in
the lattice. The average value of the BTW automaton is observed to take
different values for two different kinds of the arrangements of the dilution.
It is found that in the intermediate range of the dilution, 
for random distribution of holes, the average value
of the automaton is less than that for regular clustered distribution. Eventually, both
types of distribution provides the same value of average of automaton for
very low and high fraction of dilution.

\newpage

\setlength{\unitlength}{0.240900pt}
\ifx\plotpoint\undefined\newsavebox{\plotpoint}\fi
\sbox{\plotpoint}{\rule[-0.200pt]{0.400pt}{0.400pt}}%

\bigskip

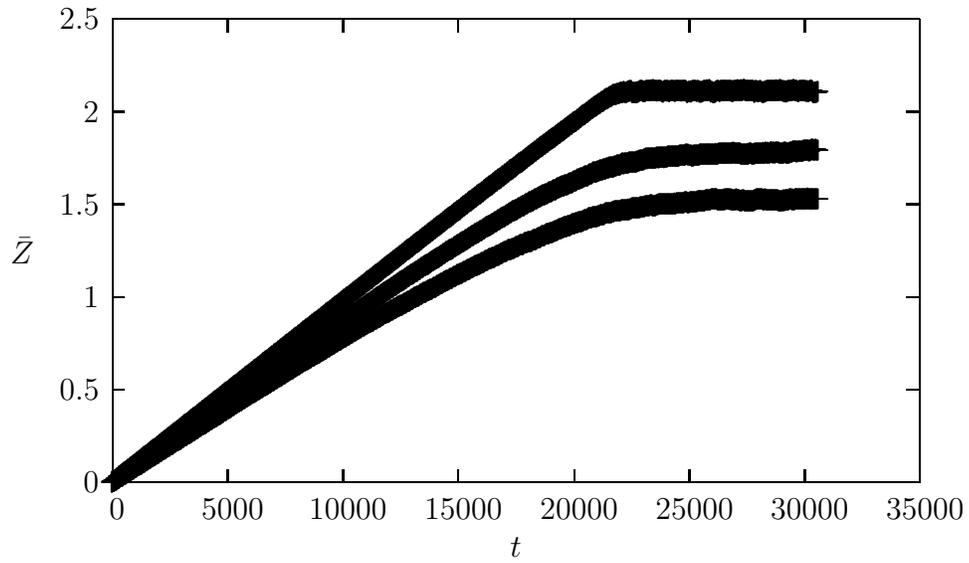
\noindent {\bf Fig. 1}. The time varitions of $\bar z$ for BTW model with 
holes (randomly scattered ) for three different values of fraction of 
holes($f_h$). 
\newpage

\setlength{\unitlength}{0.240900pt}
\ifx\plotpoint\undefined\newsavebox{\plotpoint}\fi
\sbox{\plotpoint}{\rule[-0.200pt]{0.400pt}{0.400pt}}%
%
\bigskip

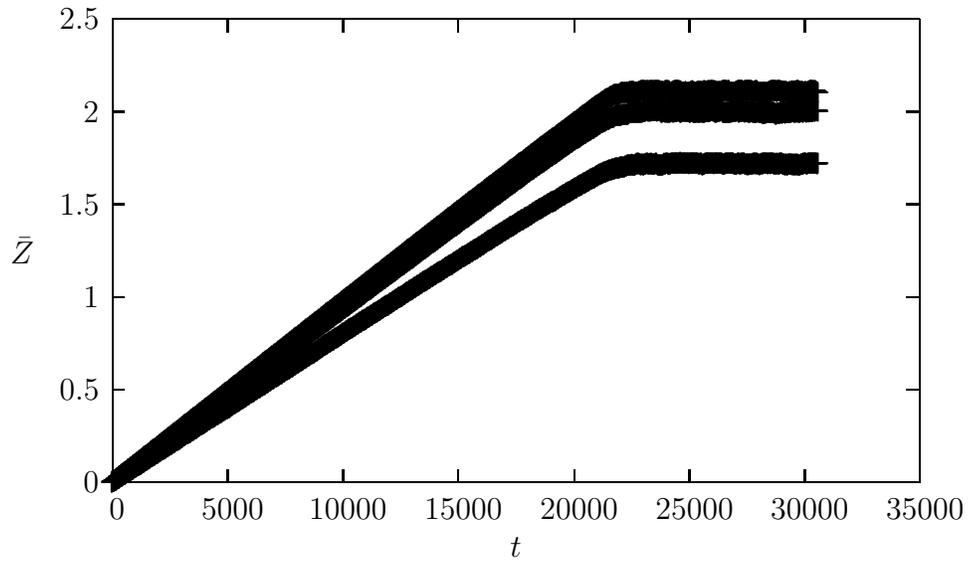
\noindent {\bf Fig. 2}. The time variations of $\bar z$ for BTW model with 
holes (as a single cluster) for three different values of the fraction of holes. 
\newpage

\setlength{\unitlength}{0.240900pt}
\ifx\plotpoint\undefined\newsavebox{\plotpoint}\fi
\sbox{\plotpoint}{\rule[-0.200pt]{0.400pt}{0.400pt}}%
%
\bigskip

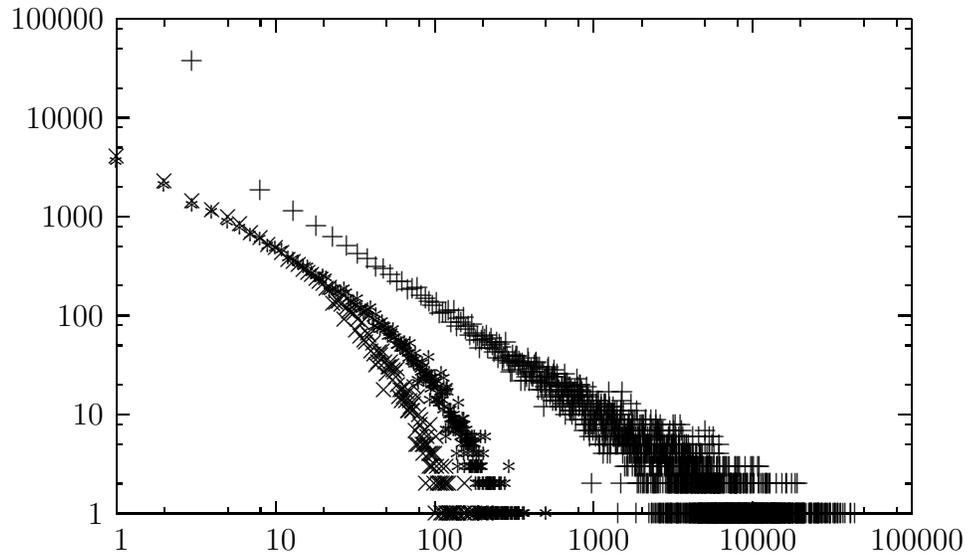
\noindent {\bf Fig. 3}. Log-Log plot of  the unnormalized distributions 
of avalanche size for BTW model and BTW model with holes (randomly scattered)  
for two different values of the  fraction of holes.

\newpage

\setlength{\unitlength}{0.240900pt}
\ifx\plotpoint\undefined\newsavebox{\plotpoint}\fi
\sbox{\plotpoint}{\rule[-0.200pt]{0.400pt}{0.400pt}}%
%
\bigskip

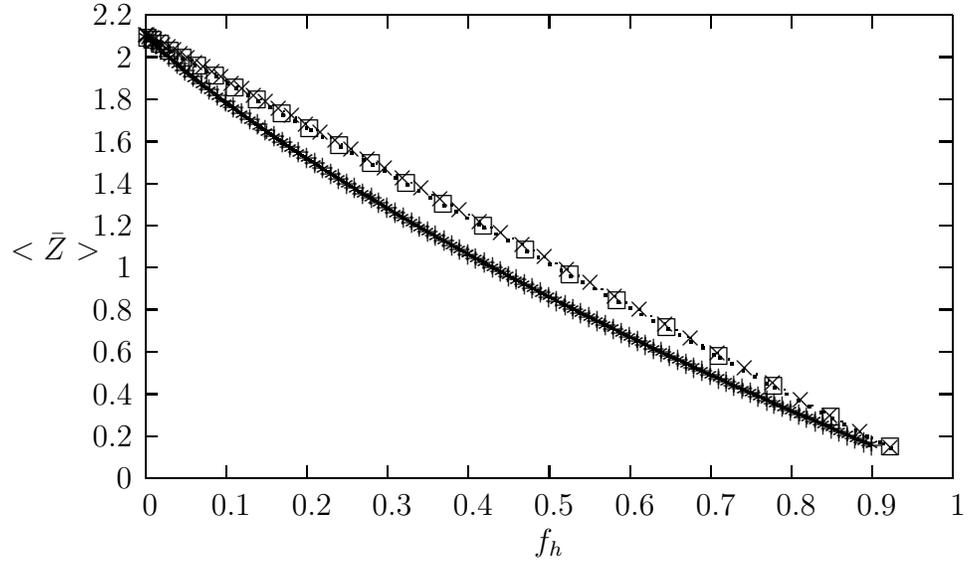
\noindent {\bf Fig. 4}. Variation of time average of $\bar z$ $(<\bar z>)$
 with fraction of holes  $(f_h)$ for BTW model with holes
when holes are randomly scattered for $L=51 (*)$ , $L=101 (+)$ 
and for a single cluster of hole for $L=51 (\Diamond)$, $L=101 (\times )$

\bigskip

\vspace {0.6 cm}
\begin{thebibliography}{99}

\bibitem{btw} P. Bak, C. Tang and K. Wiesenfeld, Phys. Rev. Lett., {\bf 59}
(1987) 381; Phys. Rev. A, {\bf 38} (1988) 364.


\bibitem{dd} D. Dhar, Phys. Rev. Lett., {\bf 64} (1990) 1613;
S. N. Majumder and D. Dhar, J. Phys. A {\bf 24} (1991) L357.


\bibitem{evi} E. V. Ivashkevich, J. Phys. A {\bf 27} (1994) 3643;
V. B. Priezzhev, J. Stat. Phys, {\bf 74} (1994) 955.

\bibitem{vbp} V. B. Priezzhev, D. V. Ktitarev and E. V. Ivashkevich,
Phys. Rev. Lett {\bf 76} (1996) 2093.

\bibitem{bb} A. Benhur and O. Biham, Phys. Rev. E {\bf 53} (1996) R1317.

\bibitem{lub} S. Lubek and K. D. Usadel, Phys. Rev. E {\bf 55} (1997) 4095.

\bibitem{ev} E. V. Ivashkevich, J. Phys. A {\bf 27 } (1994) L585.

\bibitem{piet} L. Pietronero, A. Vespignani and S. Zapperi, Phys. Rev. Lett.,
{\bf 72} (1994) 1690;
E. V. Ivashkevich, Phys. Rev. Lett, {\bf 76} (1996) 3368.

\bibitem{ab} A. Bhowal, Physica A {\bf 247} (1997) 327

\bibitem{mkk} S. S. Manna, L. B. Kiss and J. Kertesz, J. Stat. Phys., {\bf 61},
(1990) 923

\bibitem{manna2} S. S. Manna, A. D. Chakrabarti and R Cafiero, Phys. Rev. E, {\bf 60},
(1999) R5005

\bibitem{malcal} O. Malcal, Y. Shllo and O. Blham, Phy. Rev. E ,  {\bf 73}, (2006) 056125

\bibitem{najafl} M. N. Najafl, S. M. Araghl and S. Rouhanl, Phys. Rev. E, {\bf 85} (2012) 051104


\end{thebibliography}
\end{document}